\documentstyle[12pt]{article}
\begin{document}

\def\ds{\displaystyle}
\def\beq{\begin{equation}}
\def\eeq{\end{equation}}
\def\bea{\begin{eqnarray}}
\def\eea{\end{eqnarray}}
\def\beeq{\begin{eqnarray}}
\def\eeeq{\end{eqnarray}}
\def\ve{\vert}
\def\hats{\hat{s}}
\def\hatm{\hat{m}}
\def\barc{\bar{c}}
\def\Br{{\cal B}}
\def\vel{\left|}
\def\ver{\right|}
\def\nnb{\nonumber}
\def\ga{\left(}
\def\dr{\right)}
\def\aga{\left\{}
\def\adr{\right\}}
\def\lla{\left<}
\def\rra{\right>}
\def\rar{\rightarrow}
\def\nnb{\nonumber}
\def\la{\langle}
\def\ra{\rangle}
\def\ba{\begin{array}}
\def\ea{\end{array}}
\def\tr{\mbox{Tr}}
\def\ssp{{\Sigma^{*+}}}
\def\sso{{\Sigma^{*0}}}
\def\ssm{{\Sigma^{*-}}}
\def\xis0{{\Xi^{*0}}}
\def\xism{{\Xi^{*-}}}
\def\qs{\la \bar s s \ra}
\def\qu{\la \bar u u \ra}
\def\qd{\la \bar d d \ra}
\def\qq{\la \bar q q \ra}
\def\gGgG{\la g^2 G^2 \ra}
\def\q{\gamma_5 \not\!q}
\def\x{\gamma_5 \not\!x}
\def\g5{\gamma_5}
\def\sb{S_Q^{cf}}
\def\sd{S_d^{be}}
\def\su{S_u^{ad}}
\def\ss{S_s^{??}}
\def\sbp{{S}_Q^{'cf}}
\def\sdp{{S}_d^{'be}}
\def\sup{{S}_u^{'ad}}
\def\ssp{{S}_s^{'??}}
\def\sig{\sigma_{\mu \nu} \gamma_5 p^\mu q^\nu}
\def\fo{f_0(\frac{s_0}{M^2})}
\def\ffi{f_1(\frac{s_0}{M^2})}
\def\fii{f_2(\frac{s_0}{M^2})}
\def\O{{\cal O}}
\def\sl{{\Sigma^0 \Lambda}}
\def\es{\!\!\! &=& \!\!\!}
\def\ar{&+& \!\!\!}
\def\ek{&-& \!\!\!}
\def\cp{&\times& \!\!\!}
\def\se{\!\!\! &\simeq& \!\!\!}
\def\bsll{b \rar s \ell^+ \ell^-}
\def\dsp{\displaystyle}
% ...........................................................

\renewcommand{\textfraction}{0.2}    %float (figures) parameters
\renewcommand{\topfraction}{0.8}

\renewcommand{\bottomfraction}{0.4}
\renewcommand{\floatpagefraction}{0.8}
\newcommand\mysection{\setcounter{equation}{0}\section}

\def\baeq{\begin{appeq}}     \def\eaeq{\end{appeq}}
\def\baeeq{\begin{appeeq}}   \def\eaeeq{\end{appeeq}}
\newenvironment{appeq}{\beq}{\eeq}
\newenvironment{appeeq}{\beeq}{\eeeq}
\def\bAPP#1#2{
 \markright{APPENDIX #1}
 \addcontentsline{toc}{section}{Appendix #1: #2}
 \medskip
 \medskip
 \begin{center}      {\bf\LARGE Appendix #1 :}{\quad\Large\bf #2}
% \begin{center}      {\bf\LARGE Appendix  :}{\quad\Large\bf #2}
\end{center}
 \renewcommand{\thesection}{#1.\arabic{section}}
\setcounter{equation}{0}
        \renewcommand{\thehran}{#1.\arabic{hran}}
\renewenvironment{appeq}
  {  \renewcommand{\theequation}{#1.\arabic{equation}}
     \beq
  }{\eeq}
\renewenvironment{appeeq}
  {  \renewcommand{\theequation}{#1.\arabic{equation}}
     \beeq
  }{\eeeq}
\nopagebreak \noindent}

\def\eAPP{\renewcommand{\thehran}{\thesection.\arabic{hran}}}

\renewcommand{\theequation}{\arabic{equation}}
\newcounter{hran}
\renewcommand{\thehran}{\thesection.\arabic{hran}}

\def\bmini{\setcounter{hran}{\value{equation}}
\refstepcounter{hran}\setcounter{equation}{0}
\renewcommand{\theequation}{\thehran\alph{equation}}\begin{eqnarray}}
\def\bminiG#1{\setcounter{hran}{\value{equation}}
\refstepcounter{hran}\setcounter{equation}{-1}
\renewcommand{\theequation}{\thehran\alph{equation}}
\refstepcounter{equation}\label{#1}\begin{eqnarray}}

%       the stuff below defines \eqalign and \eqalignno in such a
%       way that they will run on Latex

\newskip\humongous \humongous=0pt plus 1000pt minus 1000pt
\def\caja{\mathsurround=0pt}
%\def\eqalign#1{\,\vcenter{\openup1\jot
%\caja   %\ialign{\strut \hfil$\displaystyle{##}$&$
%\displaystyle{{}##}$\hfil\crcr#1\crcr}
%}\,}

% ...........................................................

\title{
 {\Large
                 {\bf
Analysis of Various Polarization Asymmetries In The Inclusive
$b\rightarrow s \ell^+ \ell^-$ Decay In The Fourth-Generation
Standard Model
                 }
         }
      }

\author{\\
{\small V. Bashiry$^1$\thanks {e-mail: bashiry@ciu.edu.tr}, M.
Bayar$^2$\thanks {e-mail: mbayar@metu.edu.tr}\,\,, }\\ {\small $^1$
Engineering Faculty, Cyprus International University,} \\ {\small
Via Mersin 10, Turkey }\\{\small $^2$
Physics Department, Middle East Technical University,}\\
{\small 06531 Ankara, Turkey}}
\date{}
\begin{titlepage}
\maketitle
\thispagestyle{empty}

\begin{abstract}
In this study a systematical analysis of various polarization
asymmetries in inclusive $b \rar s \ell^+ \ell^-$ decay in the
standard model (SM) with four generation of quarks is carried out.
We found that the various asymmetries are sensitive to the new
mixing and quark masses for both of the $\mu$ and $\tau$ channels.
Sizeable deviations from the SM values are obtained. Hence, $b \rar
s \ell^+ \ell^-$ decay is a valuable tool for searching physics
beyond the SM, especially in the indirect searches for the
fourth-generation of quarks ($t',\, b')$.
\end{abstract}

%\vspace{1cm}
~~~PACS numbers: 12.60.--i, 13.30.--a, 14.20.Mr
\end{titlepage}

\section{Introduction}
Despite incredible successes of the Standard Model with three
generations of quarks (SM3) in explaining the experimental data, it
is believed that SM3 of electroweak interaction is a low energy
manifestation of some other more fundamental theory. Therefore,
investigations of new physics beyond the standard model (SM) is now
being performed in particle physics. A possible new physics is the
existence of new  quarks and leptons beyond the known ones. Whether
or not there exist new generations has been investigated by many
theoretical and experimental studies (for the most recent studies
see \cite{Hung,Kikukawa:2009mu} and the references therein). These
new generations might soon be detected by LHC, where the new
generation is expected to be produced by gluon fusion. The cross
section of the production of $t'\bar{t'}$ at LHC is about: $\sigma
=10(0.25)$pb for $m_{t'}=400(800)$ GeV, which is similar to the
$t\bar{t}$ production \cite{Arik:1996qd}.

The status of four generations has arisen many discussions from the
experimental point of view. On the other hand, from theoretical
point of view, it is favored because of two reasons; first, it might
help in bringing the $SU(3)\bigotimes SU(2)_L\bigotimes U(1)_Y$
couplings close to each other at the unification point $\sim10^{16}$
GeV \cite{Hung2}; second, new generations with new weak phases might
bring better solution to baryogenesis \cite{Hou:2008xd}.

The experimental constraints on the 4$^{th}$ generation are imposed
by the $\rho,\, S,\, T$ parameters of the SM and the measurement of
the Z boson decay width. LEP and CDF \cite{CDF}  experiments provide
constraints on the mass of 4$^{th}$ generation lepton (neutrinoes
heavier than half of the Z mass) and quarks (4$^{th}$ generation
quarks heavier than $255$ GeV), respectively. The LEP results
exclude the possibility of the 4$^{th}$ generation of leptons with
the mass $\sim 1$ eV \cite{PDG}. These results are mostly
interpreted as the exact value of the generation number, i.e, $N=3$,
since one assumes that the neutrinoes must have very small masses.
If we disregard this incorrect assumption, the LEP data does not
exclude the existence of extra SM families with heavy neutrinoes.
Note that, the existence of the consequential 4$^{th}$ generation
leptons besides the 4$^{th}$ generation quarks is indispensable to
cancel the contributions of the 4$^{th}$ generation quarks in the
gauge anomalies at loop level.

Flavor changing neutral current (FCNC) transition is  in forefront
of indirect investigation for the 4$^{th}$ generation via $b \rar
s(d)$ transition. These transitions, which lead to the so called
rare decays, appear at quantum level, since, they are forbidden in
tree level in the SM. The 4$^{th}$ consequential SM family of
quarks, i.e, $t'$, like $u,\,c,\, t$ quarks, can contribute to the
loop. Such consequential extension of the SM has been formulated by
many authors, i.e. \cite{Arhrib:2002md,Bashiry:2007pd,
Aliev:2003gp}. Note that, the fourth generation effects have been
widely studied in baryonic and semileptonic exclusive B decays
\cite{Hou:2006jy}--\cite{Zolfagharpour:2007eh}.  Note also that, the
calculation of the inclusive $\bsll$ decays is cleaner than the
exclusive decays because  the exclusive decays suffer from the
hadronic uncertainties. In this study, we investigate the
possibility of searching for new physics when looking at various
asymmetries in the inclusive $\bsll$ decay using the SM with fourth
generation of quarks ($b',\, t'$). Note that, branching ratio, CP
asymmetry and FB asymmetry for this decay in the fourth generation
standard model (SM4) have been studied in \cite{Arhrib:2002md}.

The paper includes 6 sections: In section 2, we modify the effective
Hamiltonian in the presence of 4$^{th}$ generation. In section 3 and
4, double lepton polarization and polarized FB asymmetries are
derived, respectively. In section 5, we examine the sensitivity of
these physical observable on the new parameters ($m_{t'}$,
$V_{t'b}V^*_{t's}$ ). Section 6 is devoted to the conclusions.
\section{Theoretical Framework}
In this section, we present the theoretical expressions for the
decay width. As we mentioned above, we extend the SM3 to the
fourth-generation standard model (SM4) and as a result of this
extension the Wilson coefficient of the SM3 is modified by the
existence of the 4$^{th}$ generation quark $t'$. It is easy to see
that if a fourth generation is introduced in the same way as the
other three generations in the SM3, new operators do not appear. In
other words, the full operator set for the SM4 is exactly the same
as in the SM3.

The Wilson coefficients are modified as follows: \bea\lambda_t C_i
\rightarrow \lambda_t C^{SM}_i+\lambda_{t'} C^{new}_i~,\eea where
$\lambda_f=V_{f b}^\ast V_{fs}$. The unitarity of the $4\times4$ CKM
matrix leads to
\bea\lambda_u+\lambda_c+\lambda_t+\lambda_{t'}=0.\eea\ One can
neglect $\lambda_u=V_{ub}^\ast V_{us}$ in Eq.~2 which is very small
in strength compared to the others ($|\lambda_u|\sim 10^{-3}$).
Then, $\lambda_t\approx -\lambda_c-\lambda_{t'}$.

 Now, we can
re-write Eq.~1 as:  \bea \lambda_t C^{SM}_i+\lambda_{t'}
C^{new}_i=-\lambda_c C^{SM}_i+\lambda_{t'} (C^{new}_i-C^{SM}_i
).\eea It is clear that when $m_{t'}\rar m_t$ or $\lambda_{t'}\rar
0$, $\lambda_{t'} (C^{new}_i-C^{SM}_i )$ vanishes, as is required by
the GIM mechanism.

The most important operators for $B\to X_s\ell^+\ell^-$ are
%%%
\bea
%%O_7 &=& \frac{\alpha_{\rm em}}{2\pi}\, {i q^\nu\over q^2}\,
%%  (\bar s_L \sigma_{\nu\mu} b_R)\, (\bar\ell\, \gamma^\mu \ell)\,, \nn\\
O_7 &=& \frac{e}{16\pi^2}\, \bar{m}_b(\mu)\,
  (\bar s_L \sigma_{\mu\nu} b_R) F^{\mu\nu} , \nnb\\
O_9 &=& \frac{\alpha_{\rm em}}{4\pi}\, (\bar s_L \gamma_\mu b_L)\,
  (\bar\ell\, \gamma^\mu \ell)\,, \nnb\\
O_{10} &=& \frac{\alpha_{\rm em}}{4\pi}\, (\bar s_L \gamma_\mu
b_L)\,
  (\bar\ell\, \gamma^\mu\gamma_5 \ell)\,.
\eea
%%%
The large $q^2$ region is usually considered less favorable, because
it has a smaller rate and suffers from large nonperturbative
corrections. However, the experimental efficiency is better in that
region \cite{Lee:2008xc}. The operator $O_7$ is dominant at small
$q^2$ due to the $1/q^2$ pole from the photon propagator.  At high
$q^2$ region the $O_7$ contribution is rather small \cite{Ligeti}.
The rate in the high $q^2$ region has a smaller renormalization
scale dependence and $m_c$ dependence~\cite{Ghinculov:2003qd}.
Despite the experimental advantages, the large $q^2$ region has been
considered less favored, because it has a large hadronic
uncertainty~\cite{Hurth:2007xa}. The $1/m_b^3$ corrections are not
much smaller than the $1/m_b^2$ ones~\cite{Bauer:1999za} when the
operator product expansion becomes an expansion in
$\Lambda_{QCD}/(m_b-\sqrt{q^2})$~\cite{Neubert:2000ch} instead of
$\Lambda_{QCD}/m_b$.

 The QCD corrected
effective Hamiltonian for the $b\rightarrow s\ell^{+}\ell^{-}$
transitions leads to the following matrix element:
\begin{eqnarray}
 M =\frac{G_{F}
V_{tb}V^{*}_{ts}}{\sqrt{2}\pi}\alpha_{em}[&C^{tot}_{9}(\overline{s}\gamma_{\mu}P_{L}b)\overline{\ell}\gamma_{\mu}\ell
+C_{10}^{tot}(\overline{s}\gamma_{\mu}P_{L}b)\overline{\ell}\gamma_{\mu}\gamma^{5}\ell
\nonumber\\&-2\,C_{7}^{tot}\overline{s}i\sigma_{\mu\nu}\frac{q^{\nu}}{q^{2}}
(m_{b}P_{R}+m_{s}P_{L})b\overline{\ell}\gamma_{\mu}\ell\,]
 , \,\,\, \label{amplitude}
\end{eqnarray}
 where  $q$ denotes the four
momentum of the lepton pair and $C_i^{tot}$'s are as follows: \bea
C_i^{tot}(\mu) &=& C_i^{eff}(\mu) + \frac{\lambda_{t'}} {\lambda_t}
C_i^{new} (\mu),\eea where the last terms in these expressions
describe the contributions of the $t^\prime$ quark to the Wilson
coefficients. $\lambda_{t'}$  can be parameterized as: \bea
{\label{parameter}} \lambda_{t'}=V_{t^\prime b}^\ast V_{t^\prime
s}=r_{sb}e^{i\phi_{sb}}.\eea
 Neglecting the terms of
$O(m_q^2/m_W^2)$, $q = u, d, c$, the analytic expressions for all
Wilson coefficients in the SM in the  leading order (LO) and in the
next to leading logarithmic approximation (NLO) can be found in
\cite{burev}--\cite{misiak}. The most recent developments in the SM
calculations of $b\rightarrow s\ell^{+}\ell^{-}$ transitions and the
results for the NNLL virtual corrections to the matrix elements of
the operators $O_1$ and $O_2$ for this inclusive process in the
kinematical region $q^2 > 4m^2_c$ have been discussed in Ref.
\cite{Ligeti,Hurth2008} and the references therein. In the
low-dilepton-mass region $q^2 < 6$ GeV$^2$ and also in the
high-dilepton-mass region with $q^2 > 4m^2_c$, theoretical
predictions for the invariant mass spectrum are dominated by the
perturbative contributions, and a theoretical precision of order
$10\%$ is in principle possible.

 The explicit forms of the $C_i^{new}$ can be obtained from the
corresponding expression of the Wilson coefficients in the SM by
substituting $m_t \rar m_{t^\prime}$.

$C_9^{eff}(\hats) = C_9 + Y(\hats)$, where $Y(\hats) = Y_{\rm
pert}(\hats) + Y_{\rm LD}$ contains both the perturbative part
$Y_{\rm pert}(\hats)$ and long-distance part $Y_{\rm LD}(\hats)$.
$Y(\hats)_{\rm pert}$ is given by \cite{Buras:1994dj}
\begin{eqnarray}
Y_{\rm pert} (\hats) &=& g(\hatm_c,\hats) C_0 \nonumber\\&&
-\frac{1}{2} g(1,\hats) (4 \bar{C}_3 + 4 \bar{C}_4 + 3 \bar{C}_5 +
\bar{C}_6) -\frac{1}{2} g(0,\hats) (\bar{C}_3 + 3 \bar{C}_4)
\nonumber\\&& +\frac{2}{9} (3 \bar C_3 + \bar C_4 + 3 \bar C_5 +
\bar C_6),
\\
\mbox{with}\quad C_0 &\equiv& \bar C_1 + 3\bar C_2 + 3 \bar C_3 +
\bar C_4 + 3 \bar C_5 + \bar C_6,
\end{eqnarray}
and the function $g(x,y)$ defined in \cite{Buras:1994dj}.  Here
$\bar{C_1}$ -- $\bar{C_6}$ are the Wilson coefficients in the
leading logarithmic approximation. The relevant Wilson coefficients
were collected in Ref. \cite{Buras:1993xp}. $Y(\hat{s})_{\rm LD}$
involves $b \to s V(\bar{c} c)$ resonances
\cite{Ali:1991is,Lim:1988yu,Kruger:1996cv}, where $V(\bar{c} c)$ are
the vector charmonium states. We follow
Refs.~\cite{Ali:1991is,Lim:1988yu} and set
\begin{eqnarray}
Y_{\rm LD}(\hats) &=&
 - \frac{3\pi}{\alpha_{em}^2} C_0
\sum_{V = \psi(1s),\cdots} \kappa_V \frac{\hatm_V \Br(V\to
\l^+\l^-)\hat{\Gamma}_{\rm tot}^V}{\hats - \hatm_V^2 + i \hatm_V
\hat{\Gamma}_{\rm tot}^V},
\end{eqnarray}
where $\hat{\Gamma}_{\rm tot}^V \equiv \Gamma_{\rm tot}^V/m_B$ and
$\kappa_V =2.3$. The relevant properties of vector charmonium states
are summarized in Table~\ref{charmonium}.
%%%%%%%%%%%%%%%%%%%%%%%%%%%%%%%%%%%%%%%%%%%%%%%%%%%%%%%%%%%%%
\begin{table}[tbp]
\caption{Masses, total decay widths and branching fractions of
dilepton decays of vector charmonium states
\cite{Yao:2006px}.}\label{charmonium}
\begin{center}
%\begin{ruledtabular}
\begin{tabular}{cclll}
$V$ & Mass[GeV] &  $\Gamma_{\rm tot}^V$[MeV]
 &\multicolumn{2}{c}{$\Br(V\to\ell^+\ell^-)$}
\\
\hline $J/\Psi(1S)$ & $3.097$ & $0.093$ & $5.9\times10^{-2}$ & for
$\l=e,\mu$
\\
$\Psi(2S)$   & $3.686$ & $0.327$ & $7.4\times10^{-3}$ & for $\l=e,\mu$ \\
             &         &             & $3.0\times10^{-3}$ & for $\l=\tau$
\\
$\Psi(3770)$ & $3.772$ & $25.2$ & $9.8\times10^{-6}$ & for $\l=e$
\\
$\Psi(4040)$ & $4.040$ & $80$ & $1.1\times10^{-5}$ & for $\l=e$
\\
$\Psi(4160)$ & $4.153$ & $103$ & $8.1\times10^{-6}$ & for $\l=e$
\\
$\Psi(4415)$ & $4.421$ & $62$ & $9.4\times10^{-6}$ & for $\l=e$
%\\
\end{tabular}
%\end{ruledtabular}
\end{center}
\end{table}

 Using the expression of matrix element in
equation (\ref{amplitude}) and neglecting the s-quark mass
($m_s$)~\cite{aali}--\cite{falk}, we obtain the expression for the
differential decay rate as~\cite{babu}:
\begin{eqnarray}
\Gamma_0\,=\frac{d \Gamma}{d \hat{s}} = \frac{G_F m_b^5}{192
\pi^3} \frac{\alpha_{em}^2}{4 \pi^2} |V_{tb} V_{ts}^*|^2 (1 -
\hat{s})^2 \sqrt{1 - \frac{4 \hat{m}_{\ell}^2}{\hat{s}}}
\bigtriangleup \label{difdecaywidth}
\end{eqnarray}
 with
\begin{eqnarray}
\bigtriangleup &=& 4 \frac{(2 + \hat{s})}{\hat{s}} \left(1 +
\frac{2 \hat{m}_{\ell}^2}{\hat{s}}\right) |C_7^{tot}|^2 + (1 + 2
\hat{s}) \left(1 + \frac{2 \hat{m}_{\ell}^2}{\hat{s}}\right)
|C_9^{tot}|^2 \nonumber \\
&& + (1 - 8 \hat{m}_{\ell}^2 + 2 \hat{s} + \frac{2
\hat{m}_{\ell}^2}{\hat{s}}) |C_{10}^{tot}|^2  + 12 (1 + \frac{2
\hat{m}_{\ell}^2}{\hat{s}}) Re(C_9^{tot *} C_7^{tot})  .
\label{delta}
\end{eqnarray}
\section{Polarization Asymmetries}
In order to compute the polarization asymmetries, one has to choose
a reference frame to define the spin directions. A reference frame
can be chosen in the center of mass (CM) frame of the leptons. In
such a reference frame, if we suppose that $\ell^-$ moves in
positive $z$ direction and the fact that momentum is conserved, the
$s$ and $b$ quarks move in the same direction. In this reference
frame, the spin direction of leptons, the 4-vector $s^\mu_{\ell^-}$,
after the Lorentz boost from its rest frame can be obtain as
\cite{Bensalem:2002ni}:

\beq s^\mu_{\ell^-} = \left\{ {|p^-|\over m_\ell} s^-_z , s^{-}_x ,
s^{-}_y, {E \over m_\ell} s^{-}_z \right\} ~~,~~~~ s^\mu_{\ell^+} =
\left\{ {|p^+|\over m_\ell} s^+_z , s^+_x , s^{+}_y, {E \over
m_\ell} s^{+}_z \right\} ~, \label{frame} \eeq

where  $\bf s^{\pm}$ and $p^{\pm}$ are the unit vectors and
three-momenta of leptons in the $\ell^{\pm}$ rest frames,
respectively. %%
The double--lepton polarization asymmetries ${\cal P}_{ij}$ are
defined as \cite{Fukae}%%

\beq {\cal P}_{ij} =\frac{\Big[\frac{d\Gamma({\bf s^{+}}={\bf
\hat{i}},{\bf
      s^{-}}={\bf \hat{j}})}{d\hat{s}}-\frac{d\Gamma({\bf s^{+}}={\bf
      \hat{i}},{\bf s^{-}}={-\bf \hat{j}})}{d\hat{s}}\Big]
  -\Big[\frac{d\Gamma({\bf s^{+}}={-\bf \hat{i}},{\bf s^{-}}={\bf
      \hat{j}})}{d\hat{s}}-\frac{d\Gamma({\bf s^{+}}={-\bf
      \hat{i}},{\bf s^{-}}={-\bf \hat{j}})}{d\hat{s}}\Big]}
{\Big[\frac{d\Gamma({\bf s^{+}}={\bf \hat{i}},{\bf s^{-}}={\bf
      \hat{j}})}{d\hat{s}}+\frac{d\Gamma({\bf s^{+}}={\bf
      \hat{i}},{\bf s^{-}}={-\bf \hat{j}})}{d\hat{s}}\Big]
  +\Big[\frac{d\Gamma({\bf s^{+}}={-\bf \hat{i}},{\bf s^{-}}={\bf
      \hat{j}})}{d\hat{s}}+\frac{d\Gamma({\bf s^{+}}={-\bf
      \hat{i}},{\bf s^{-}}={-\bf \hat{j}})}{d\hat{s}}\Big]},
\eeq
where ${\hat i}$ and ${\hat j}$ are unit vectors.

With our choice of reference frame Eq.~(\ref{frame}), the decay
happens in two dimensions, i.e., $yz$ plane. In this frame, just the
components of  spin can be in $\hat x$ direction. Therefore, any
terms including the spin along $\hat x$ direction are the result of
either dot product of two spins or triple-product correlation with
one spin along $\hat x$ direction (i.e.,\ ${\cal P}_{xx}$, ${\cal
P}_{xy}$ and ${\cal P}_{xz}$). This holds even in the presence of
any extension of SM. Among these quantities, ${\cal P}_{xy}$ and
${\cal P}_{xz}$ are interesting as they probe the imaginary parts of
the products of Wilson coefficients \cite{Bensalem:2002ni}.

The ${\cal P}$'s take the form
\begin{eqnarray} \label{2}
{\cal P}_{xx}&=&\frac{1}{\Delta} \Bigg
\{24Re(C_{7}^{tot}C_{9}^{tot^{\ast}})\frac{\hat{m}{_{\ell}^{2}}}{\hat{s}}
+4|C_{7}^{tot}|^{2}\frac{(-1+\hat{s})\hat{s}+2(2+\hat{s})\hat{m}{_{\ell}^{2}}}{\hat{s}^{2}}\nonumber
\\&&+(|C_{9}^{tot}|^{2}-|C_{10}^{tot}|^{2})\frac{(1-\hat{s})\hat{s}+2(1+2\hat{s})\hat{m}{_{\ell}^{2}}}{\hat{s}}\Bigg \},\\
{\cal P}_{yx}&=&\frac{-2}{\Delta}\
Im(C_{9}^{tot}C_{10}^{tot^{\ast}})(1-\hat{s})\sqrt{1-\frac{4\hat{m}{_{\ell}^{2}}}{\hat{s}}},\\
{\cal P}_{xy}&=&{\cal P}_{yx},\\
{\cal
P}_{zx}&=&\frac{-3\pi}{2\sqrt{\hat{s}}\Delta}\hat{m}{_{\ell}}\
\Bigg\{2Im(C_{7}^{tot}C_{10}^{tot^{\ast}})+Im(C_{9}^{tot}C_{10}^{tot^{\ast}})\Bigg\},\\
{\cal
P}_{yy}&=&\frac{1}{\Delta}\Bigg\{24Re(C_{7}^{tot}C_{9}^{tot^{\ast}})
\frac{\hat{m}{_{\ell}^{2}}}{\hat{s}}-4(|C_{9}^{tot}|^{2}+
|C_{10}^{tot}|^{2})\frac{(1-\hat{s})\hat{m}{_{\ell}^{2}}}{\hat{s}}\nonumber
\\&&+(|C_{9}^{tot}|^{2}-|C_{10}^{tot}|^{2})((-1+\hat{s})+\frac{6\hat{m}{_{\ell}^{2}}}{\hat{s}})\nnb
\\&&+4|C_{7}^{tot}|^{2}\frac{((1-\hat{s})\hat{s}
+2(2+\hat{s})\hat{m}{_{\ell}^{2}})}{\hat{s}^{2}}\Bigg\},\\
{\cal P}_{zy}&=&\frac{3\pi}{2\sqrt{\hat{s}}\Delta}\
\Bigg\{2Re(C_{7}^{tot}C_{10}^{tot^{\ast}})-|C_{10}^{tot}|^{2}+Re(C_{9}^{tot}C_{10}^{tot^{\ast}})\hat{s}\Bigg\}\hat{m}{_{\ell}}
\sqrt{1-\frac{4\hat{m}{_{\ell}^{2}}}{\hat{s}}},\\
{\cal P}_{xz}&=&-{\cal P}_{zx},\\
{\cal P}_{yz}&=&\frac{3\pi}{2\sqrt{\hat{s}}\Delta}\
\Bigg\{2Re(C_{7}^{tot}C_{10}^{tot^{\ast}})+|C_{10}^{tot}|^{2}
+Re(C_{9}^{tot}C_{10}^{tot^{\ast}})\hat{s}\Bigg\}\hat{m}{_{\ell}}
\sqrt{1-\frac{4\hat{m}{_{\ell}^{2}}}{\hat{s}}},\\
{\cal
P}_{zz}&=&\frac{1}{2\Delta}\Bigg\{12Re(C_{7}^{tot}C_{9}^{tot^{\ast}})
(1-\frac{2\hat{m}{_{\ell}^{2}}}{\hat{s}})
+\frac{4|C_{7}^{tot}|^{2}(2+\hat{s})(1-\frac{2\hat{m}{_{\ell}^{2}}}{\hat{s}})}{\hat{s}}\nonumber
\\&&+(|C_{9}^{tot}|^{2}+|C_{10}^{tot}|^{2})(1+2\hat{s}-\frac{6(1+\hat{s})\hat{m}{_{\ell}^{2}}}{\hat{s}})\nonumber
\\&&+\frac{2(|C_{9}^{tot}|^{2}-|C_{10}^{tot}|^{2})(2+\hat{s})\hat{m}{_{\ell}^{2}}}{\hat{s}}\Bigg\}.
\end{eqnarray}

Except ${\cal P}_{zz}$ which is two times smaller than the one
obtained in Ref.~\cite{Bensalem:2002ni}, the other coefficients
${\cal P}_{ij}$'s can be obtained from results in
Ref.~\cite{Bensalem:2002ni} by the replacement of
$C_i^{tot}\rightarrow C_i^{eff}$ where $i=7, ~9, ~10$.

\section{Single and Double Lepton Polarization Forward-Backward Asymmetries}
Equipped with the definition of the spin directions in the CM frame
of leptons, we can evaluate the forward-backward asymmetries
corresponding to various polarization components of the $\ell^-$
and/or $\ell^+$ spin by writing \cite{Bensalem:2002ni}:
\bea
  A_{FB}({\bf s^{+}},{\bf s^{-}},\hat{s})&=& A_{FB}(\hat{s})+ \Big[
{\cal A}^{-}_x s_x^- +{\cal A}^{-}_y s_y^- +{\cal A}^{-}_z s_z^-
+{\cal A}^{+}_x s_x^+ +{\cal A}^{+}_y s_y^+ +{\cal A}^{+}_z s_z^+
\nnb
\\
&& \hskip1truein +~{\cal A}_{xx} s_x^+ s_x^- +{\cal A}_{xy} s_x^+
                s_y^- +{\cal A}_{xz} s_x^+ s_z^- \nnb\\
&& \hskip1truein +~{\cal A}_{yx} s_y^+ s_x^- +{\cal A}_{yy} s_y^+
                s_y^- +{\cal A}_{yz} s_y^+ s_z^- \nnb \\
&& \hskip1truein +~{\cal A}_{zx} s_z^+ s_x^- +{\cal A}_{zy} s_z^+
                s_y^- +{\cal A}_{zz} s_z^+ s_z^- \Big] ~.
\eea
The different polarized forward-backward asymmetries are then
calculated as follows:
\bea
{\cal A}^{+}_x &=& 0, \\
{\cal A}^{+}_y &=& \frac{2}{\Delta}\,{{\rm Re}(C_9^{tot}
     C_{10}^{tot^*})}\,\frac{(1-\hat{s})\,
     \hat{m}_\ell}{\sqrt{\hat{s}}}\,\sqrt{1 -
     \frac{4\,\hat{m}_\ell^2}{\hat{s}}}, \\
{\cal A}^{+}_z &=&\frac{1}{\Delta}\Bigg \{ 6\,{{\rm Re}(C_7^{tot}
     C_9^{tot^*})} - \frac{6\,|C_7^{tot}|^2}{\hat{s}} -
     3\,(\,|C_9^{tot}|^2-|C_{10}^{tot}|^2)\,\hat{m}_\ell^2 \nnb\\ && -
     12\,{{\rm Re}(C_7^{tot} C_{10}^{tot^*})}\,\frac{\hat{m}_\ell^2}
     {\hat{s}} - ~6\,{{\rm Re}(C_9^{tot}
     C_{10}^{tot^*})}\,\frac{\hat{m}_\ell^2} {\hat{s}} \nnb \\ &&
     -~\frac{3}{2}\,(\,|C_9^{tot}|^2+|C_{10}^{tot}|^2)\,\hat{s}\, ( 1 -
     \frac{2\,\hat{m}_\ell^2}{\hat{s}})\Bigg \}, \\
{\cal A}^{-}_x &=& 0,\\
{\cal A}^{-}_y &=& {\cal A}^{+}_y, \\
{\cal A}^{-}_z &=&\frac{1}{\Delta}\Bigg \{ -6\,{{\rm Re}(C_7^{tot}
     C_9^{tot^*})} - \frac{6\,|C_7^{tot}|^2}{\hat{s}} -
     3\,(\,|C_9^{tot}|^2-|C_{10}^{tot}|^2)\,\hat{m}_\ell^2 \nnb\\ &&
     +~12\,{{\rm Re}(C_7^{tot} C_{10}^{tot^*})}\,\frac{\hat{m}_\ell^2}
     {\hat{s}} +~6\,{{\rm Re}(C_9^{tot}
     C_{10}^{tot^*})}\,\frac{\hat{m}_\ell^2} {\hat{s}} \nnb \\ &&
     -~\frac{3}{2}\,(\,|C_9^{tot}|^2+|C_{10}^{tot}|^2)\,\hat{s}\, ( 1 -
     \frac{2\,\hat{m}_\ell^2}{\hat{s}})\Bigg \}, \\
{\cal A}_{xx} &=& 0,\\
{\cal A}_{xy} &=& \frac{-6}{\Delta}\,( 2\,{{\rm
Im}(C_7^{tot}\,C_{10}^{tot^*})} + {{\rm
Im}(C_9^{tot}\,C_{10}^{tot^*})})\,\frac{{{\hat{m}_\ell}}^2}{\hat{s}}, \\
{\cal A}_{xz} &=& \frac{2}{\Delta}\,{{\rm
     Im}(C_9^{tot}C_{10}^{tot^*})}\,\frac{(1-\hat{s})\,{\hat{m}_\ell}
     }{\sqrt{\hat{s}}}\,\sqrt{1 -
     \frac{4\,\hat{m}_\ell^2}{\hat{s}}},
     \\
{\cal A}_{yx} &=& -{\cal A}_{xy},\\
{\cal A}_{yy} &=& 0,\\
{\cal A}_{yz}\! &=& \!\Big( 2
      |C_9^{tot}|^2-\frac{8\,|C_7^{tot}|^2}{\hat{s}} \Big) \, \frac{(
      1 - \hat{s})\,\hat{m}_\ell}{\Delta\sqrt{\hat{s}}}, \\
{\cal A}_{zx} &=& {\cal A}_{xz},\\
{\cal A}_{zy} &=& {\cal A}_{yz}, \\
{\cal A}_{zz} &=& \frac{-3}{\Delta}\,( 2\,{{\rm
     Re}(C_7^{tot}\,C_{10}^{tot^*})} + {{\rm
     Re}(C_9^{tot}\,C_{10}^{tot^*})}\,\hat{s}) \, {\sqrt{1 -
     \frac{4\,{{\hat{m}_\ell}}^2}{\hat{s}}}} ~.
\eea

Here, ${\cal A}_{zz}$ coincides with $-{\cal A}_{FB}$ in the SM and
consequential extension of the SM (SM4)
\cite{Bensalem:2002ni,Bashiry:2007tt}. A significant difference
between ${\cal A}_{zz}$ and $({\cal A}_{FB})$ appears when the new
type of interactions are taken into account in the effective
Hamiltonian, i.e, the tensor type and scalar type interactions
differ between  ${\cal A}_{zz}$ and $({\cal A}_{FB}) $(see ref.
\cite{Aliev:2004hi}).

Note that,  ${\cal A}_{ij}$ coefficients calculated in
Ref.~\cite{Bensalem:2002ni} can again be obtained by the replacement
$C_i^{tot}\rightarrow C_i^{eff}$ where $i=7, ~9, ~10$.

 \section{Numerical analysis}
We try to analyze the dependency of the various asymmetries on the
fourth generation quark mass ($m_{t'}$) and the product of quark
mixing matrix elements ($V_{t^\prime b}^\ast V_{t^\prime
s}=r_{sb}e^{i\phi_{sb}}$).
 We will use the next--to--leading order
logarithmic approximation for the Wilson coefficients $C_i^{eff}$
and $C^{new}_i$ ~\cite{Buras:1993xp,R46216} at the scale
$\mu=m_b=4.8~GeV$. It is worth to mention that, beside the short
distance contribution, $C_9^{eff}$ has also long distance
contributions resulting from the real $\bar c c$ resonant states of
the $J/\psi$ family. In the present study, we do  take the long
distance effects into account by using the approachs of
Refs.~\cite{Ali:1991is,Lim:1988yu}

The input parameters we used in this
analysis are as follows:\\
$\vel V_{tb} V_{ts}^\ast\ver = 0.04166$, $m_t=175~GeV$ ,
$m_W=80.41~GeV$ and $\Gamma_B = 4.22\times 10^{-13}~GeV$.
 In order to perform quantitative analysis of the physical
 observables, numerical values for
the new parameters ($m_{t'},\,r_{sb},\,\phi_{sb}$) are necessary.
Using the experimental values of $B\rar X_s \gamma$ and $B\rar X_s
\ell^+ \ell^-$, the bound on $r_{sb}\sim\{0.01-0.03\}$ has been
obtained \cite{Arhrib:2002md,Zolfagharpour:2007eh} for
$\phi_{sb}\sim\{0-2\pi\}$ and $m_{t'}\sim\{200,600\}~$(GeV)(see
table 2). Also considering $\Delta m_{B_s}$, $\phi_{sb}$ receives a
strong restriction ($\phi_{sb}\sim\pi/2$) \cite{Hou:2006jy}.
\begin{table}
\renewcommand{\arraystretch}{1.5}
\addtolength{\arraycolsep}{3pt}
$$
\begin{array}{|c|c |c|}
\hline  r_{sb} & 0.01 &0.02 \\
\hline
m_{t'}(GeV) &529 & 385 \\
\hline
\end{array}
$$
\caption{The  experimental limit of $ m_{t'} $ for
$\phi_{sb}=\pi/3$\cite{Zolfagharpour:2007eh}}
\renewcommand{\arraystretch}{1}
\addtolength{\arraycolsep}{-3pt}
\end{table}

\begin{table}
\renewcommand{\arraystretch}{1.5}
\addtolength{\arraycolsep}{3pt}
$$
\begin{array}{|c|c |c|}
\hline  r_{sb} & 0.01 &0.02 \\
\hline
m_{t'}(GeV) &373 & 289 \\
\hline
\end{array}
$$
\caption{The experimental limit of $ m_{t'} $ for
$\phi_{sb}=\pi/2$\cite{Zolfagharpour:2007eh}}
\renewcommand{\arraystretch}{1}
\addtolength{\arraycolsep}{-3pt}
\end{table}

In order to do simplify analysis of the observables, we must
eliminate some of the variables.  From explicit expressions of the
various physical observables, we see that they depend on four
variables $m_{t'},\,r_{sb},\,\phi_{sb}$ and $\hat s$. Therefore, it
may experimentally be difficult to study these dependencies at the
same time. For this reason, we will do two types of analysis: first,
we choose fixed values for $m_{t'}=400~GeV,\,r_{sb}\sim\{0.01,
0.02\}$ and $\phi_{sb}\sim\{\pi/3,\pi/2\}$ and look at the $\hat s $
dependency of the FB asymmetries. Note that, zero point position of
the FB asymmetries in terms of the $\hat s$ is less sensitive to the
hadronic uncertainties in exclusive decay channels. Second, we
eliminate the $\hat s$ dependence by performing integration over
$\hat{s}$ in the allowed region, i.e., we consider the averaged
values of the various asymmetries. The average gained over $\hat{s}$
is defined as: \bea \la {\cal{P(A)}} \ra = \frac{\ds \int_{4
\hat{m}_\ell^2}^{(1-\sqrt{\hat{r}_K})^2} {\cal{P(A)}} \frac{d{\cal
B}}{d \hat{s}} d \hat{s}} {\ds \int_{4
\hat{m}_\ell^2}^{(1-\sqrt{\hat{r}_K})^2} \frac{d{\cal B}}{d \hat{s}}
d \hat{s}}~.\nnb \eea We analysis the uncertainties among the SM
parameters namely, product of quarks mixing angles
$V_{tb}V^{*}_{ts}$ and quarks mass ($m_b$ and $m_c$). With present
bound on $0.03966<|V_{tb}V^{*}_{ts}|<0.04166$ \cite{Yao:2006px} we
find that the uncertainties of lepton polarization and FB
asymmetries are negligible (less than $0.1\%$). There are rather
weak dependency on the $m_c/m_b$ which we show those by relevant
figures. We present figures for only those observables that has
significant dependence on the new parameters. The extra
signs($+,~\times,~\Box,~\times\!\!\!\!\!\!\!=$) in figures show the
experimental limit on $m_{t'}$, considering the $1\sigma$ level
deviation from the measured branching ratio of $B\rightarrow X_s
\ell^-\ell^+$(see Table 1,2). {\it Note that, ${\cal
A}_{z}^-,\,{\cal A}_{xz}={\cal A}_{zx},\, {\cal A}_{xy}=-{\cal
A}_{yx},\,{\cal A}_{yz}={\cal A}_{zy}$ and ${\cal P}_{zz},\,{\cal
P}_{xz}$ for $\mu$ channel and ${\cal A}_{y}^+={\cal
A}_{y}^-,\,{\cal A}_{xz}={\cal A}_{zx},\, {\cal A}_{yz}=-{\cal
A}_{zy},\,{\cal P}_{yx}={\cal P}_{yx}$ and ${\cal P}_{zz}$ for
$\tau$ channel do not deviate from the SM3 values over than $10\%$.
Hence, we do not present their predictions in the figures.} A
typical deviation of order $-5\%$ to $-10\%$ for studying the ratio
of physical observables such as asymmetries, as we can see from
figures and above mentioned discussion, can not be covered by the
uncertainties among the SM parameters \cite{Lee:2008xc} i.e., the
uncertainties of quark quark masses, quarks mixing angles and higher
order calculations of the Wilson coefficients. Here, large parts of
the uncertainties partially cancel out.

From these figures, we deduce the following results:

\subsection{ Differential Polarized FB Asymmetries}
Figs 1--3 depict the $\hat s$ dependency of the single or double
lepton polarization FB asymmetries for fixed value of the 4$^{th}$
generation quark mass ($m_{t'}=400$~GeV) and different values of the
$r_{sb}$ and $\phi_{sb}$.
\begin{itemize}

\item{${\cal A}_z^-(\hat s)$  for $\tau$
channel strongly depend on the SM4 parameters. The discrepancy of
${\cal A}_z^-(\hat s)$ with respect to the $m_c/m_b$ is almost
negligible (see figs. 1).}

\item{Magnitude of ${\cal A}_{zz}(\hat s)$ is suppressed by the
4$^{th}$ generation for $\mu$ channel (see figs. 2). The zero point
position of ${\cal A}_{zz}(\hat s)$ for $\mu$ channel stays the same
as the SM3 case. This point is especially important for the
exclusive decays where the hadronic uncertainty almost vanishes at
this point. The deviation is considerable for low and high $\hat s$
values where both are in the non--resonance region (see fig. 2).}

\item{${\cal A}_{xy}(\hat s)=-{\cal A}_{yx}(\hat s)$ strongly
depends on SM4 parameters. The deviation can be a high as a factor
of seven in the low $\hat s$ region. At high $\hat s$ region the
sensitivity is less than the low $\hat s$ region (see fig. 3).
Moreover, there is rather weak dependency on the value of
$m_c/m_b$.}
\end{itemize}

\subsection{ Averaged Double Lepton Polarization Asymmetries}
\begin{itemize}
\item{ Taking into account the 4$^{th}$ generation, the value of
$\la{\cal P}_{xx} \ra$, $\la{\cal P}_{yz}\ra$ and $\la{\cal P}_{yy}
\ra$ show strong dependency on the new parameters for both $\mu$ and
$\tau$ channels.  While $\la{\cal P}_{xx} \ra$, $\la{\cal
P}_{yz}\ra$ and $\la{\cal P}_{yy} \ra$ are increasing for ,
$\phi_{sb}=\pi/2$, they increase/decrease in different regions of
parameter space for $\phi_{sb}=\pi/3$ (see figs. 4--8). The
dependency on the value of $m_c/m_b$ is ignorable.
     }

\item{ Due to inclusion of 4$^{th}$ generation, the value of $\la{\cal
P}_{xz} \ra$ gets sizable deviation from the SM3 value (which is
almost zero) (see fig. 9). Compared to the SM3 prediction, the
$\tau$ channel obtains the maximum value  about -0.2 (-0.3) when
$\phi_{sb}\sim{\pi/2(\pi/3)}$ (see fig. 9). Also, there is weak
dependency on the value of $m_c/m_b$.
     }
\item{ The non--zero values of $\la{\cal P}_{yx}\ra$ in the SM3
has their origin in the higher order QCD corrections to the
$C_9^{eff}$. Since this function is proportional to imaginary part
of the $C_9^{eff}$, its value is negligible. But, it exceeds the SM3
value sizeably with the SM4 contribution. This is because of the new
weak phase and new contribution to the Wilson coefficients coming
from the 4$^{th}$ generation. Furthermore, the maximum value of
$\la{\cal P}_{yx}\ra$ for $\mu$ channel is almost independent from
the values of $r_{sb}$ and depend on $\phi_{sb}$ (see fig. 10). Note
that its value also is sensitive to the value of $m_c/m_b$.}
\end{itemize}

Finally, the quantitative estimate about the accessibility of the
various physical observables in experiments are in order. To observe
an asymmetry A at the $n \sigma$ level, the required number of $B
\bar{B}$ pairs is given as: \bea N = \frac{n^2}{{\cal B} s_1 s_2 \la
{\cal A} \ra^2}~,\nnb \eea where $s_i (i=1,2)$ is the efficiency  of
the lepton and ${\cal B}$ is the branching ratio.

Typical values of the efficiencies of the $\tau$--leptons range from
$50\%$ to $90\%$ for their various decay modes \cite{R6016}. It
should be noted, here, that the error in $\tau$--lepton polarization
is estimated to be about $(10 \div 15)\%$ \cite{R6017}. So, the
error in measurement of the $\tau$--lepton asymmetries is
approximately $(20 \div 30)\%$, and the error in obtaining the
number of events is about $50\%$.
 Equipped with the
expression of  $N$, it can be understood that in order to detect
the asymmetries in the $\mu$ and $\tau$ channels at $3\sigma$
level with the asymmetry of ${\cal A}=10\%$ and efficiency of
$\tau \sim 0.5$), the minimum number of required events are $
N\sim 10^8$ and $N\sim 10^9$ for $\mu$ and $\tau$ leptons,
respectively.

On the other hand, the number of $B \bar{B}$ pairs, that are
produced at  LHC, are about $\sim 10^{12}$. As a result of
comparison of these numbers and $N$, we conclude that a typical
asymmetry of (${\cal A}=10\%$) is detectable at LHC.

\section{Conclusion}
To sum up, we present the various asymmetries in inclusive $b \rar s
\ell^+ \ell^-$ transition using the SM with the 4$^{th}$ generation
of quarks. The results are:
\begin{itemize}

\item{The zero point position of the polarized single or double
lepton polarization FB asymmetry coincide with each other in the SM3
and SM4. Furthermore, there are sizable discrepancies, specially, in
the non--resonance region between the result of the SM3 and SM4. }

 \item{ Some of the
double--lepton polarization and polarized double or single lepton
polarization Forward--Backward asymmetries which are already
accessible at LHC depict the strong dependency on the 4$^{th}$
generation quark mass and product of quark mixing.}

\item{While the magnitude of asymmetries, which is proportional to
the real part of the product of Wilson coefficients, is generally
suppressed by the 4$^{th}$ generation parameters. The situation for
the asymmetries proportional to the imaginary part of the product of
Wilson coefficients is enhanced}

\item{We examine the uncertainties among the SM parameters and we show that the discrepancy of order
$10\%$  in principle can not be covered by the uncertainties among
the input parameters.}

Thus, the study of such strong dependent asymmetries can serve as
good test for the predictions of the SM3 and indirect search for
the 4th generation up type quarks $t'$.

\end{itemize}

\section{Acknowledgment}
The authors would like to thank T. M. Aliev and A. Ozpineci for
their useful discussions. One of the authors (M.B.) would like to
thank TUBITAK, Turkish Scientific and Research Council, for their
financial support provided under the project 103T666. Also, V.
Bashiry would like to thank Ministry of Education of North Cyprus
(Turkish Republic), for their partial financial support through the
project.

\newpage

\section*{Figure captions}

%{\bf Fig. (1)} The dependence of the $ {\cal A}_y^+(\hat s)=-{\cal
%A}_y^-(\hat s)$ for the $b \rar s \mu^+ \mu^-$ decay on $\hat s$ for
%the fourth generation quark mass $m_{t'}=400$~GeV for two different
%values of
% $\phi_{sb}=60^\circ,~ 90^\circ$ and $r_{sb}=0.01,~0.02$ .\\ \\
%{\bf Fig. (2)} The same as in Fig. (1), but for the $ {\cal
%A}_z^+(\hat
%s)$ .\\ \\
%{\bf Fig. (3)} The same as in Fig. (2), but for the $\tau$ lepton.\\ \\
{\bf Fig. (1)} The dependence of the ${\cal A}_z^-(\hat s)$ for the
$b \rar s \tau^+ \tau^-$ decay on $\hat s$ for the fourth generation
quark mass $m_{t'}=400$~GeV for two different values of
 $\phi_{sb}=60^\circ,~ 90^\circ$ and $r_{sb}=0.01,~0.02$ .\\ \\
 {\bf Fig. (2)} The dependence of the $ {\cal A}_{zz}(\hat s)$ for the $b \rar s \mu^+ \mu^-$ decay on $\hat s$ for
the fourth generation quark mass $m_{t'}=400$~GeV for two different
values of
 $\phi_{sb}=60^\circ,~ 90^\circ$ and $r_{sb}=0.01,~0.02$ .\\ \\
 %{\bf Fig. (6)} The same as in Fig. (5), but for the $\tau$ lepton .\\ \\
 {\bf Fig. (3)} The dependence of the ${\cal A}_{xy}(\hat s)=-{\cal A}_{yx}(\hat s)$ for
the $b \rar s \tau^+ \tau^-$ decay on $\hat s$ for the fourth
generation quark mass $m_{t'}=400$~GeV for two different values of
 $\phi_{sb}=60^\circ,~ 90^\circ$ and $r_{sb}=0.01,~0.02$ .\\ \\
%{\bf Fig. (8)} The same as in Fig. (5), but for the ${\cal A}_{xz}(\hat s)$.\\ \\
%{\bf Fig. (9)} The dependence of the $\la{\cal A}_{zz}\ra$ on the
%fourth generation quark mass $m_{t'}$ for two different values of
% $\phi_{sb}=60^\circ,~ 90^\circ$ and $r_{sb}=0.01,~0.02$ for $\mu$ lepton.\\ \\
{\bf Fig. (4)} The dependence of the $\la{\cal P}_{xx}\ra$ on the
fourth generation quark mass $m_{t'}$ for two different values of
 $\phi_{sb}=60^\circ,~ 90^\circ$ and $r_{sb}=0.01,~0.02$ for $\mu$ lepton.\\ \\
 {\bf Fig. (5)} The same as in Fig. (4), but for the $\tau$ lepton.\\ \\
%{\bf Fig. (12)} The same as in Fig. (10), but for the $\la{\cal P}_{yz}\ra$.\\ \\
 {\bf Fig. (6)} The same as in Fig. (5), but for the $\la{\cal P}_{yz}\ra$.\\ \\
{\bf Fig. (7)} The dependence of the $\la{\cal P}_{yy}\ra$ on the
fourth generation quark mass $m_{t'}$ for two different values of
 $\phi_{sb}=60^\circ,~ 90^\circ$ and $r_{sb}=0.01,~0.02$ for $\mu$ lepton.\\ \\
{\bf Fig. (8)} The same as in Fig. (7), but for the $\tau$ lepton.\\ \\
 {\bf Fig. (9)} The same as in Fig. (8), but for the $\la{\cal P}_{xz}\ra$.\\ \\
{\bf Fig. (10)} The same as in Fig. (7), but for the $\la{\cal P}_{yx}\ra$.\\ \\

%\newpage
%\begin{figure} \vskip 1.5 cm
 %   \special{psfile=aypspim4.eps hoffset=-40 voffset=-270 hscale=110 vscale=110 angle=0}
%\vskip 7.8cm \caption{}
%\end{figure}

%\begin{figure}
%\vskip 2.5 cm
 %   \special{psfile=azpspim4.eps hoffset=-40 voffset=-270 hscale=110 vscale=110 angle=0}
%\vskip 7.8 cm \caption{}
%\end{figure}
%
%\begin{figure}
%\vskip 1.5 cm
 %   \special{psfile=azpspim4t.eps hoffset=-40 voffset=-220 hscale=110 vscale=110 angle=0}
%\vskip 5.8cm \caption{}
%\end{figure}

\begin{figure}
\vskip 2.5 cm
    \includegraphics{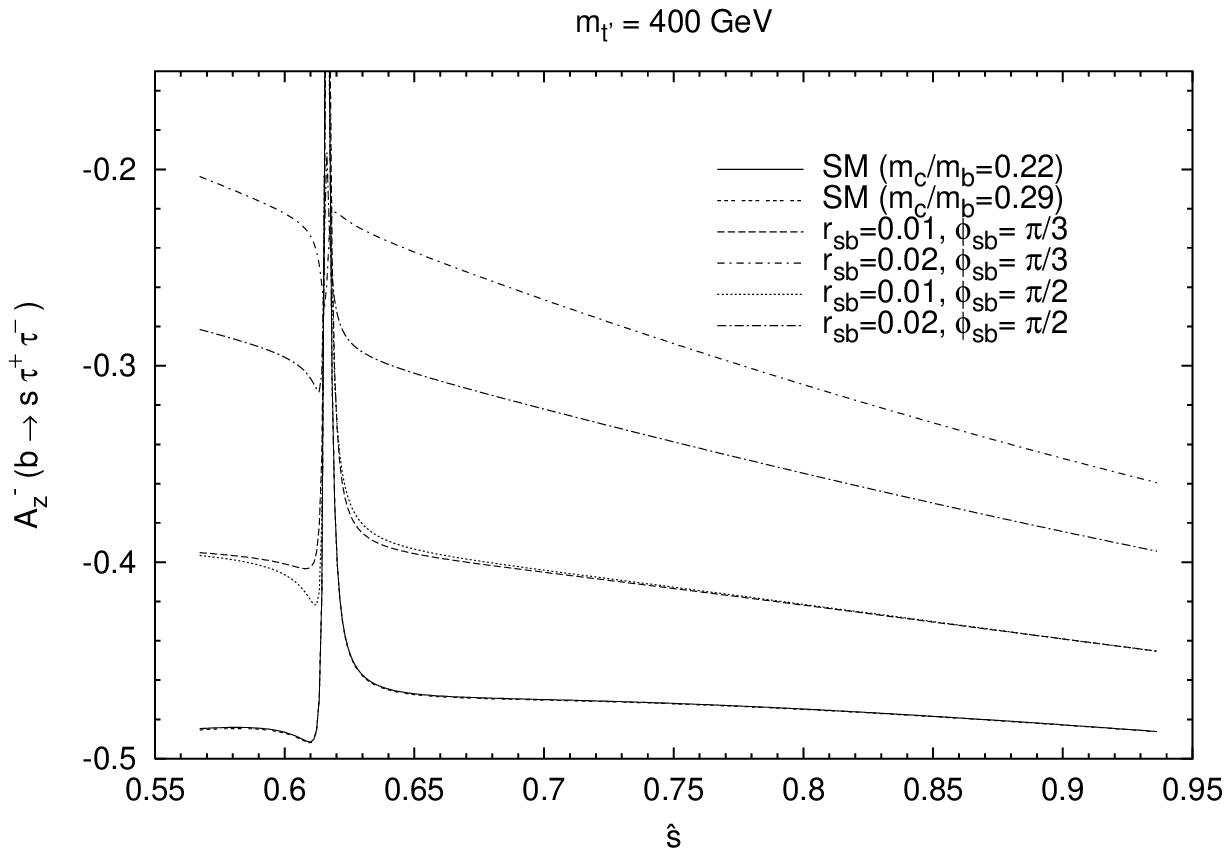}
\vskip 5.8 cm \caption{}
\end{figure}

\begin{figure}
\vskip 1.5 cm
    \includegraphics{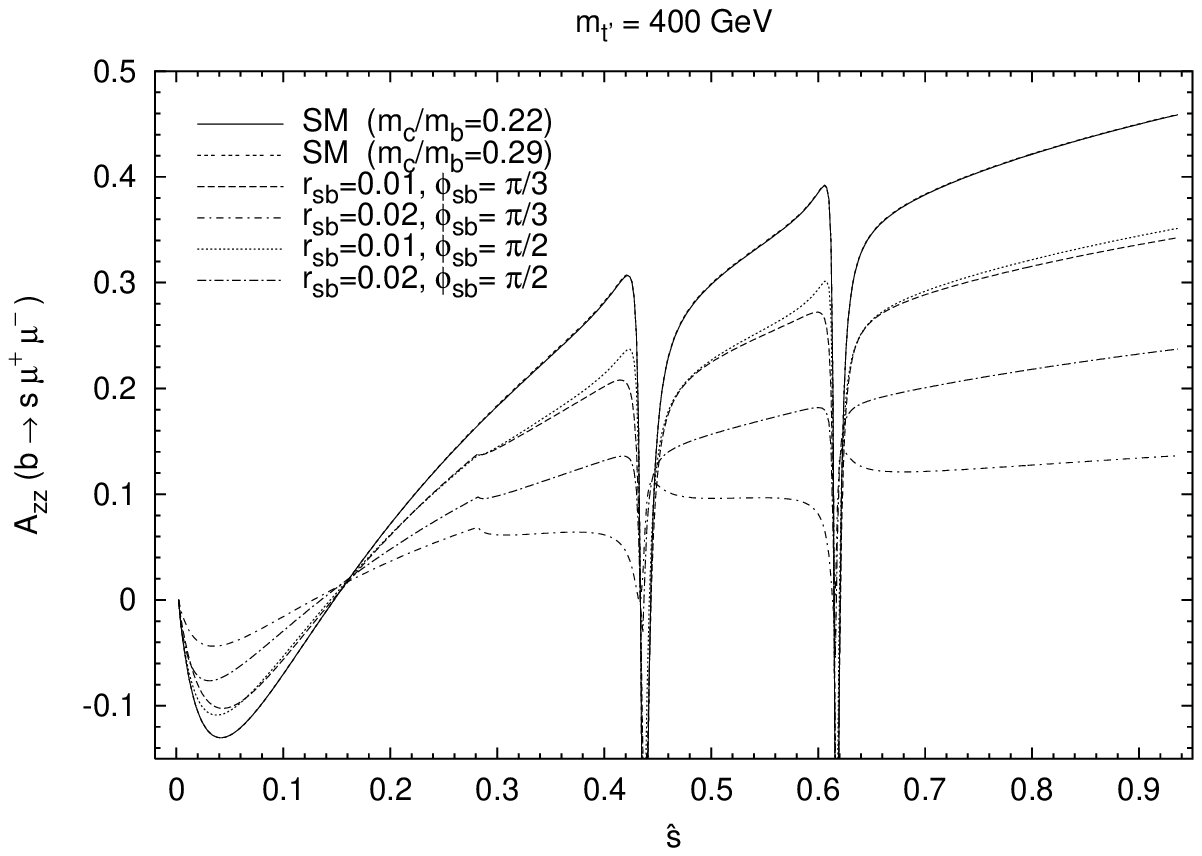}
\vskip 7.8cm \caption{}
\end{figure}

%\begin{figure}
%\vskip 2.5 cm
  %  \special{psfile=azzspim4t.eps hoffset=-40 voffset=-270 hscale=110 vscale=110 angle=0}
%\vskip 7.8 cm \caption{}
%\end{figure}

\begin{figure}
\vskip 1.5 cm
    \includegraphics{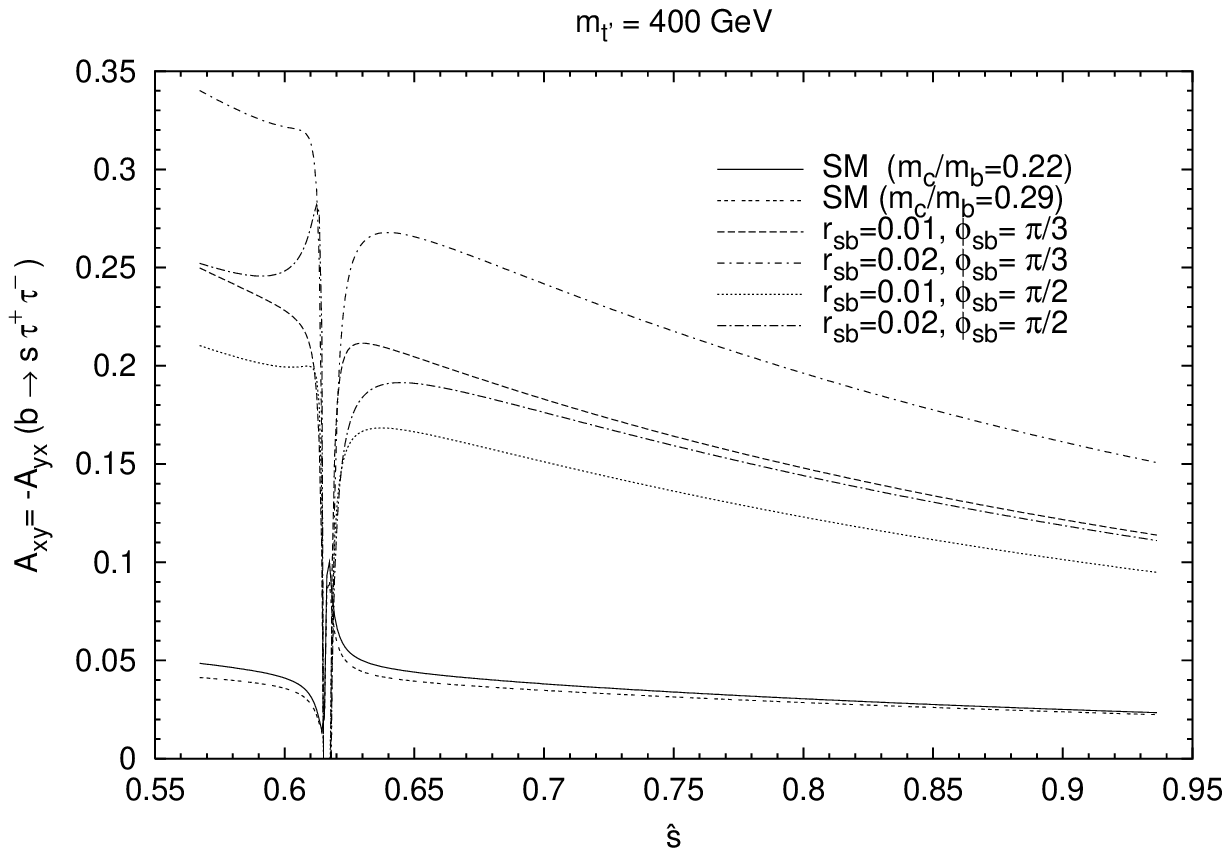}
\vskip 7.8cm \caption{}
\end{figure}

%\begin{figure}
%\vskip 2.5 cm
 %   \special{psfile=axzspim4.eps hoffset=-40 voffset=-270 hscale=110 vscale=110 angle=0}
%\vskip 7.8 cm \caption{}
%\end{figure}

%\begin{figure}
%\vskip 1.5 cm
%    \special{psfile=azpspi.eps hoffset=-40 voffset=-270 hscale=110 vscale=110 angle=0}
%\vskip 7.8cm \caption{}
%\end{figure}

%\begin{figure}
%\vskip 2.5 cm
%    \special{psfile=azmspit.eps hoffset=-40 voffset=-270 hscale=110 vscale=110 angle=0}
%\vskip 7.8 cm \caption{}
%\end{figure}

%\begin{figure}
%\vskip 1.5 cm
 %   \special{psfile=azzspi.eps hoffset=-40 voffset=-270 hscale=110 vscale=110 angle=0}
%\vskip 7.8cm \caption{}
%\end{figure}

%\begin{figure}
%\vskip 2.5 cm
%    \special{psfile=azzspit.eps hoffset=-40 voffset=-270 hscale=110 vscale=110 angle=0}
%\vskip 7.8 cm \caption{}
%\end{figure}

%\begin{figure}
%\vskip 1.5 cm
%    \special{psfile=axyspit.eps hoffset=-40 voffset=-270 hscale=110 vscale=110 angle=0}
%\vskip 7.8cm \caption{}
%\end{figure}
%
%\begin{figure}
%\vskip 2.5 cm
%    \special{psfile=ayxspit.eps hoffset=-40 voffset=-270 hscale=110 vscale=110 angle=0}
%\vskip 7.8 cm \caption{}
%\end{figure}

\begin{figure}
\vskip 1.5 cm
    \includegraphics{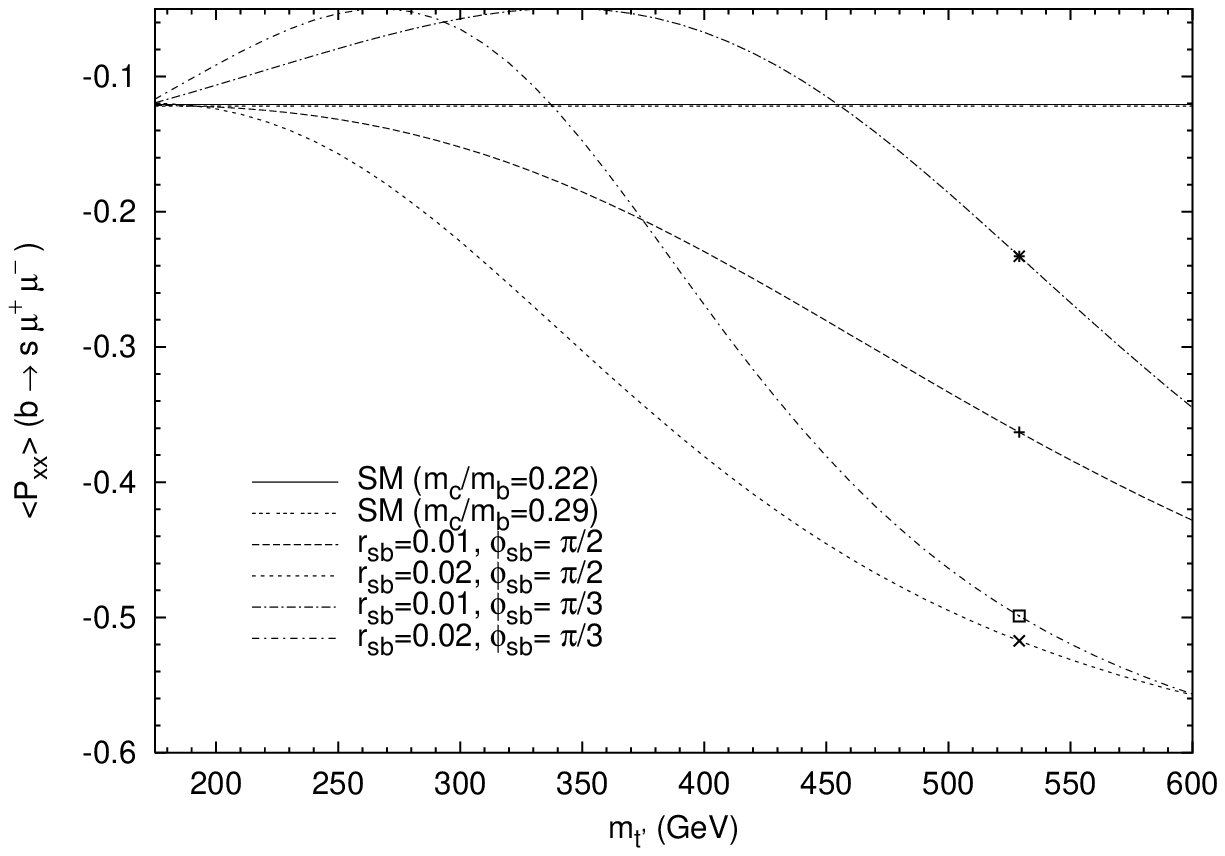}
\vskip 7.8cm \caption{}
\end{figure}

\begin{figure}
\vskip 2.5 cm
    \includegraphics{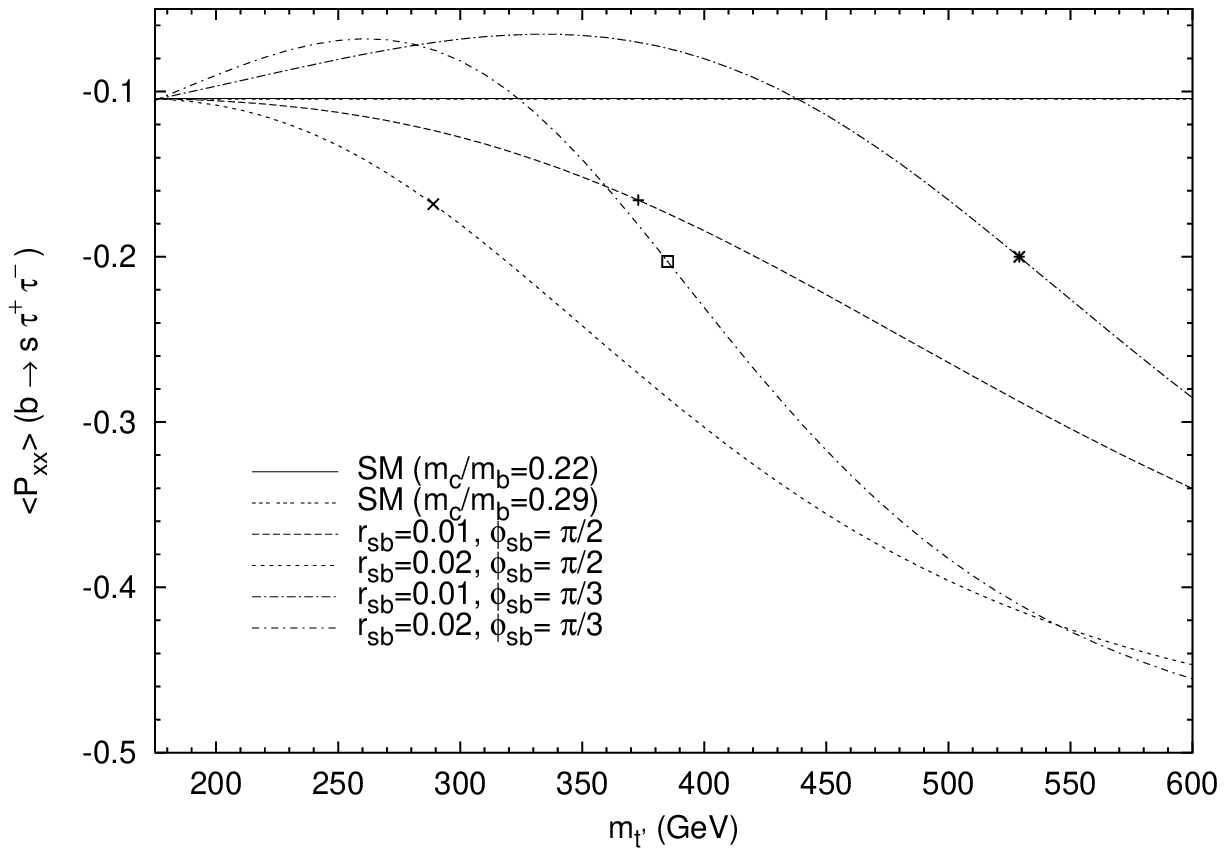}
\vskip 7.8 cm \caption{}
\end{figure}

%\begin{figure} \vskip 1.5 cm
 %   \special{psfile=pyzpi.eps hoffset=-40 voffset=-270 hscale=110 vscale=110 angle=0}
%\vskip 7.8cm \caption{}
%\end{figure}

\begin{figure}
\vskip 2.5 cm
    \includegraphics{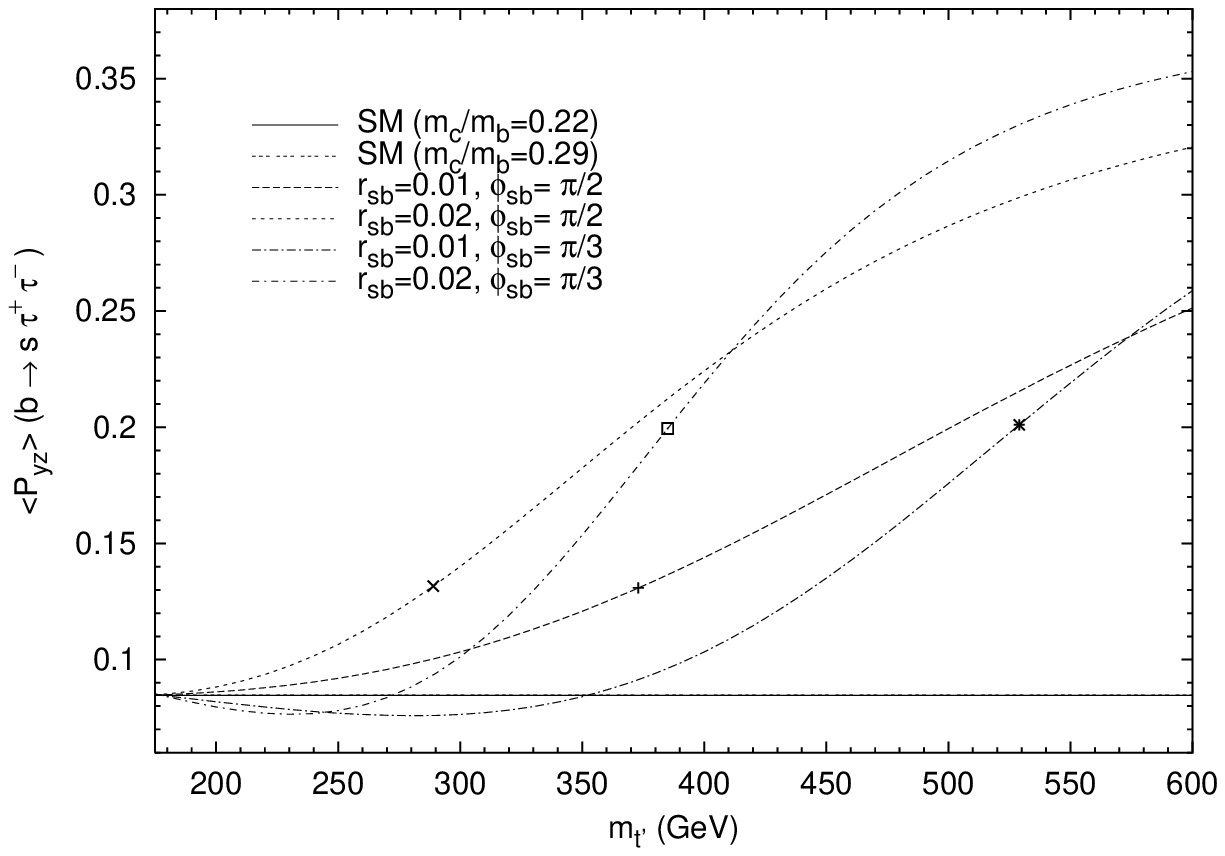}
\vskip 7.8 cm \caption{}
\end{figure}

%\begin{figure}
%\vskip 1.5 cm
%    \special{psfile=pzypi.eps hoffset=-40 voffset=-270 hscale=110 vscale=110 angle=0}
%\vskip 7.8cm \caption{}
%\end{figure}

%\begin{figure}
%\vskip 2.5 cm
%    \special{psfile=pzypit.eps hoffset=-40 voffset=-270 hscale=110 vscale=110 angle=0}
%\vskip 7.8 cm \caption{}
%\end{figure}

\begin{figure}
\vskip 1.5 cm
    \includegraphics{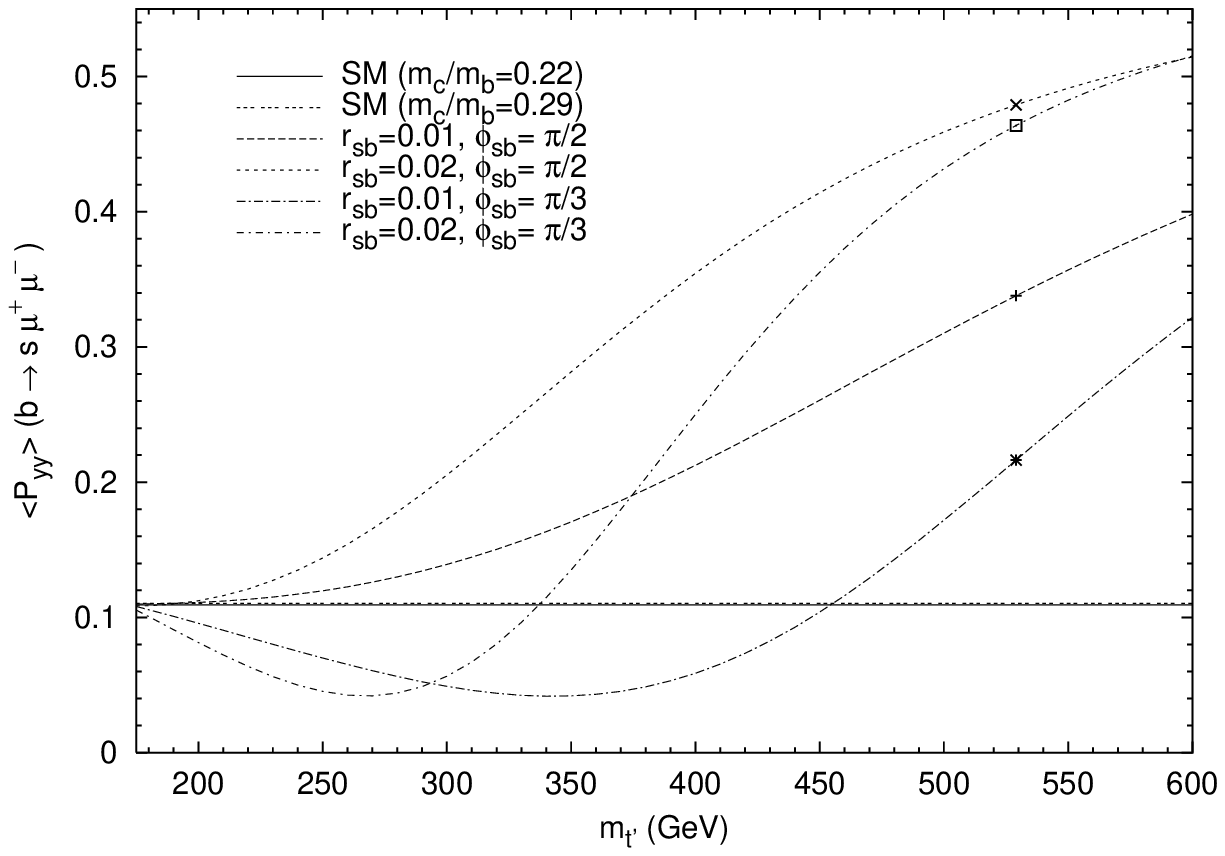}
\vskip 7.8cm \caption{}
\end{figure}

\begin{figure}
\vskip 2.5 cm
    \includegraphics{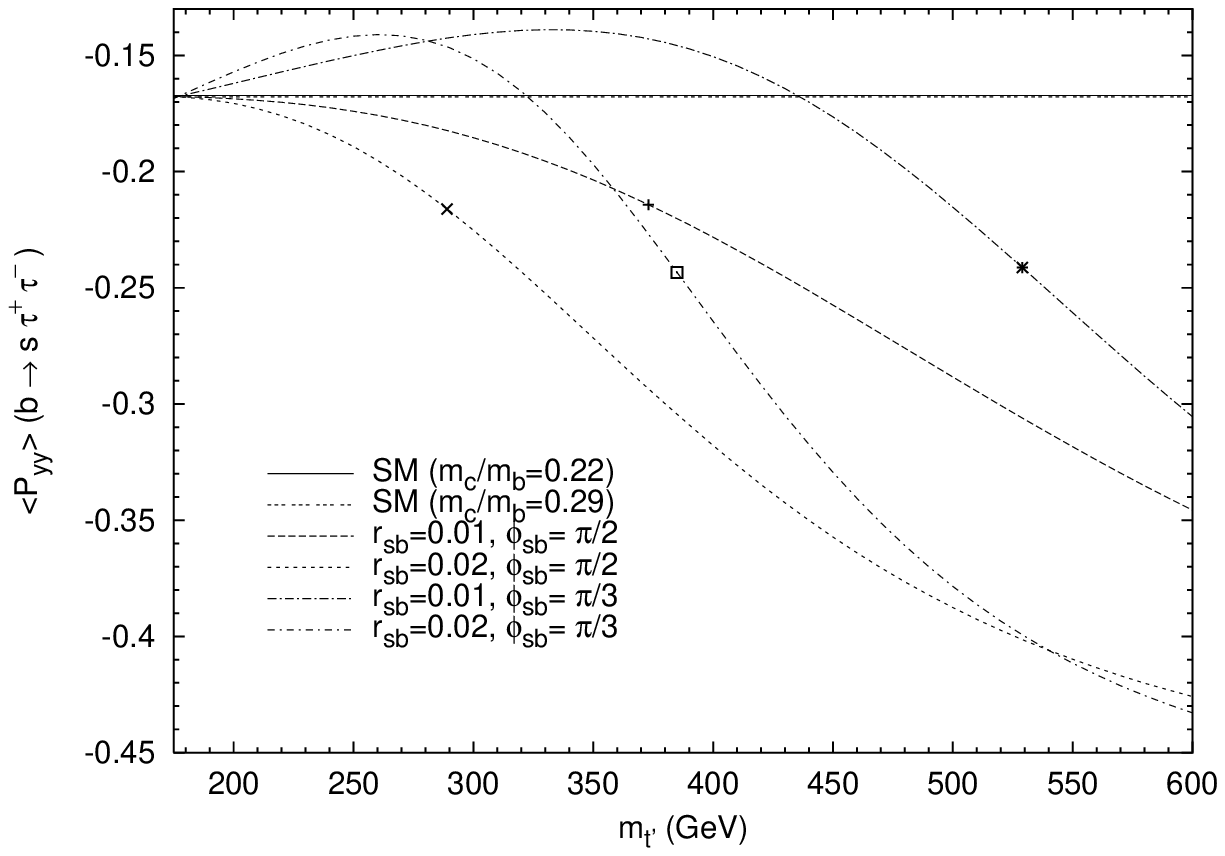}
\vskip 7.8 cm \caption{}
\end{figure}

%\begin{figure}
%\vskip 1.5 cm
%    \special{psfile=pxzpi.eps hoffset=-40 voffset=-270 hscale=110 vscale=110 angle=0}
%\vskip 7.8cm \caption{}
%\end{figure}

\begin{figure}
\vskip 2.5 cm
    \includegraphics{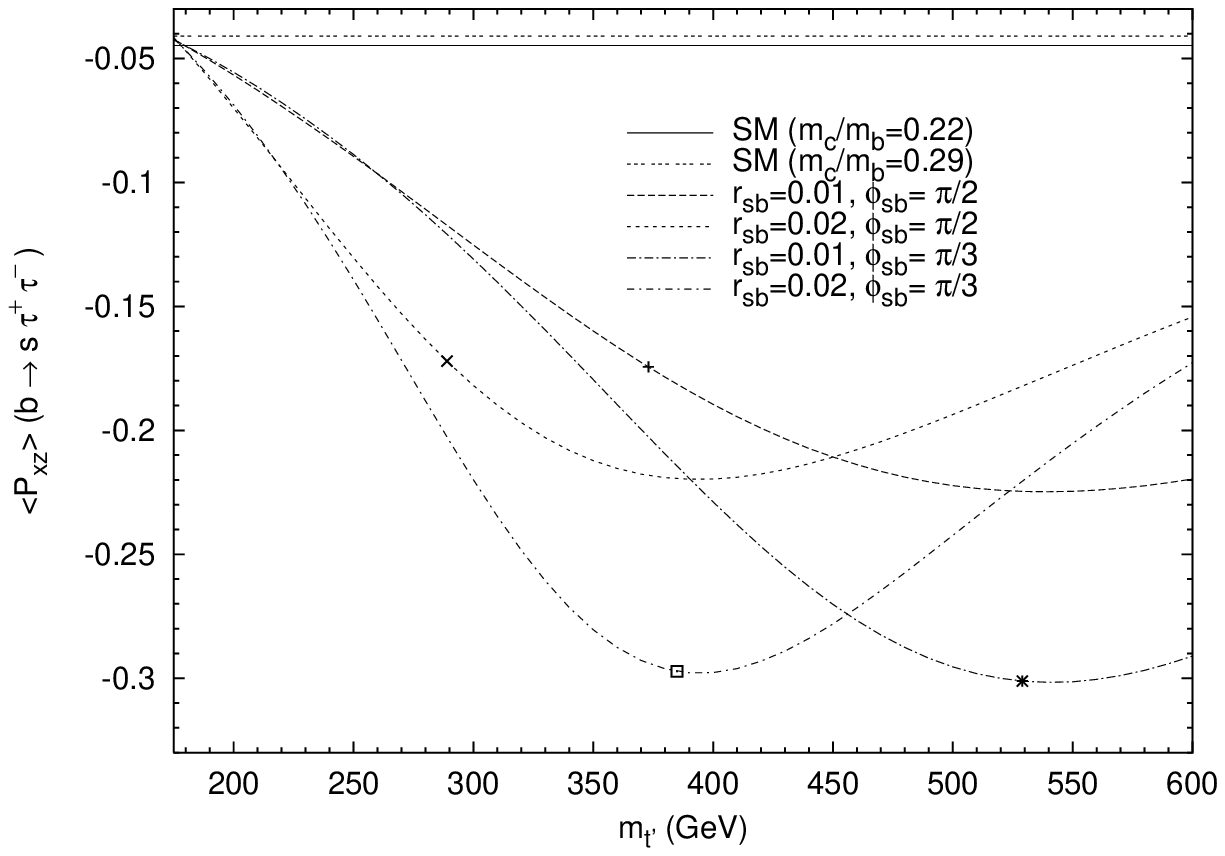}
\vskip 7.8 cm \caption{}
\end{figure}

\begin{figure} \vskip 1.5 cm
    \includegraphics{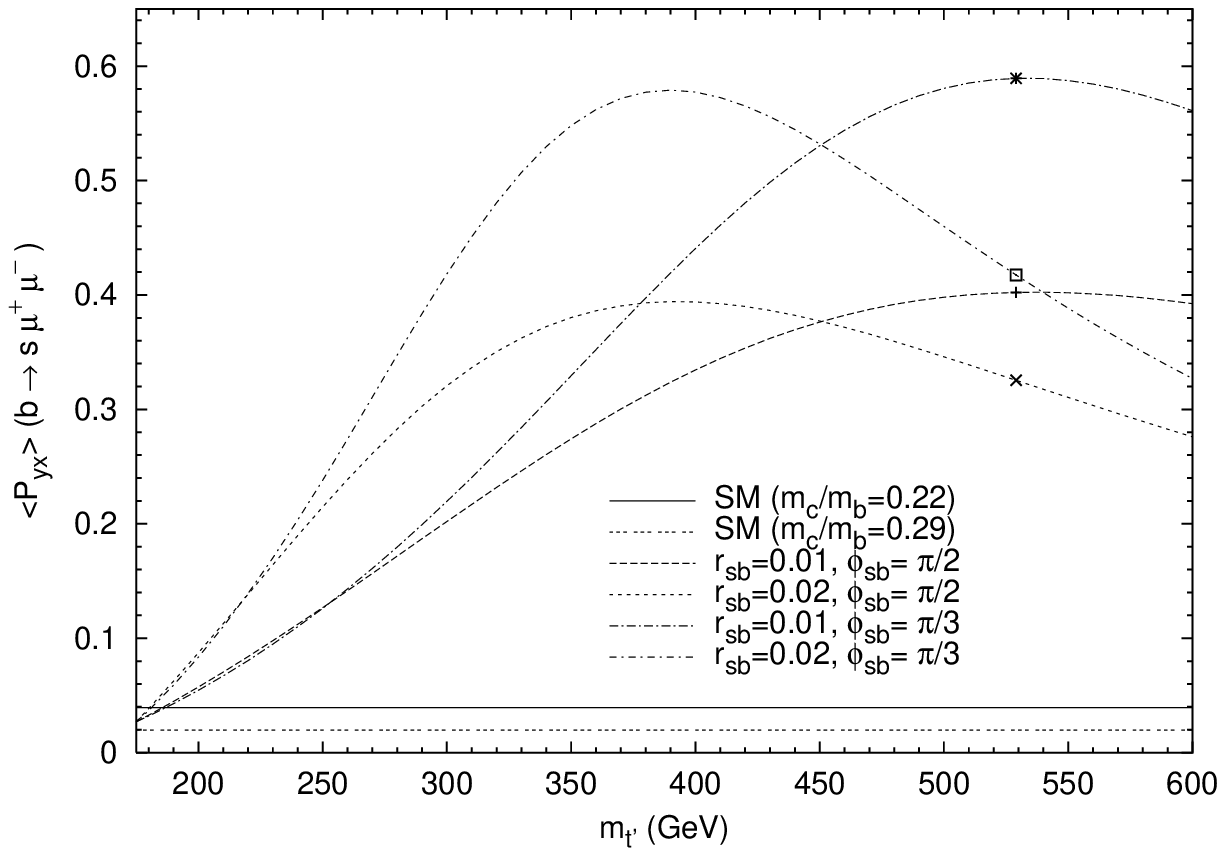}
\vskip 7.8cm \caption{}
\end{figure}

\end{document}